\documentclass[12pt]{article}

\usepackage{setspace}
\usepackage{graphicx}
\usepackage{multirow}
\usepackage{amsmath,amssymb,amsfonts}
\usepackage{amsthm}
\usepackage{mathrsfs}
\usepackage{appendix}
\usepackage{xcolor}
\usepackage{textcomp}
\usepackage{booktabs}
\usepackage{algorithm}
\usepackage{algorithmicx}
\usepackage{algpseudocode}
\usepackage{listings}
\usepackage{authblk}

\usepackage{natbib}

\definecolor{darkblue}{RGB}{24,45,158}
\usepackage[colorlinks=true,citecolor=darkblue,linkcolor=blue,urlcolor=darkblue]{hyperref}

\title{\vspace{-15mm}\fontsize{24pt}{10pt}\selectfont
\textbf{Bayesian Species Distribution Models using Hierarchical Decomposition Priors}}

\author{
\begin{minipage}[t]{0.45\textwidth}
\centering
Luisa Ferrari\\
Department of Economics\\
University of Modena and Reggio Emilia
\end{minipage}
\hfill
\begin{minipage}[t]{0.45\textwidth}
\centering
Massimo Ventrucci\\
Department of Statistical Sciences \\
University of Bologna
\end{minipage}

\vspace{2mm}

\begin{minipage}[t]{0.45\textwidth}
\centering
Alex Laini\\
Department of Life Sciences and Systems Biology\\
University of Turin
\end{minipage}
}
\date{}

\begin{document}

\maketitle

\begin{abstract}

Understanding the relative contributions of environmental, spatial, and temporal processes in shaping species distribution is a central objective in ecology. Bayesian species distribution models (SDMs) offer a flexible framework for this task, yet prior specification for variance components remains challenging.

To address this issue, we adapt the Hierarchical Decomposition (HD) prior framework to latent Gaussian SDMs, enabling direct and transparent prior control over variance partitioning. The HD approach reparametrizes variances into a total variance and a set of interpretable proportions, structured through a decomposition tree that reflects both model architecture and ecologically meaningful groupings of effects.

We discuss a principled approach for a default tree design tailored to SDMs and a practical workflow for the step-by-step implementation of the method. The framework is illustrated using presence--absence data for 39 demersal fish species from the NOAA Northeast Fisheries Science Center fall bottom trawl survey. Results demonstrate predictive performance comparable to established priors, while providing substantially improved interpretability and transparency in variance attribution and prior sensitivity analysis.


\end{abstract}

\section{Introduction}

Species distribution models (SDM) are widely used in ecology  to model species occurrence. This type of model assumes that the distribution of species' occurrence, over an area of interest and a given period of time, is the result of the combination of 3 different factors \citep{warton2015so,ovaskainen2017make}: abiotic factors, i.e. environmental characteristics; biotic factors, i.e. within and between-species interactions (e.g. predation, competition, and mutualism); additional stochastic processes (e.g. ecological drift or environmental stochasticity). Disentangling the relative contribution of these factors remains a key objective in conservation, even though a fully robust statistical framework for doing so has yet to be established.

Formally, SDMs consider a response $Y$, capturing occurrence of a species, recorded in terms of either presence-absence, count of individuals, or percentage estimate of biomass. The location and time of observation ($\boldsymbol{Z}=[Z_1,Z_2],T$) are usually available, along with a set of environmental covariates ($\boldsymbol{X}=[X_1,\dots,X_P]$), collected because assumed to be associated with occurrence.

Assume $Y$ is distributed according to an exponential family distribution with linear predictor $\eta$; a generic SDM can be written as:
\begin{gather}\label{eq:SDM}
    \eta =\mu +\sum_{j=1}^J f_j(\boldsymbol{X})+\sum_{l=1}^L f_l(\boldsymbol{Z},T).
\end{gather}
This model class is able to accommodate main effects for each of the covariates $X_1,\dots,X_P$, e.g. $f_1(X_1) + \cdots + f_P(X_P) + f_{\boldsymbol{Z}}(\boldsymbol{Z}) + f_T(T)$, but also interactions between the different environmental variables, a spatio-temporal interaction, additional unstructured noise, etc. 
The $f_j(\cdot)$ and the $f_l(\cdot)$ functions, which represent the effect of abiotic and biotic factors, can be conveniently approximated choosing to adopt the Bayesian Latent Gaussian Model (LGM) framework \citep{rue2009inla,BHM_ref}:
\begin{subequations}\label{eq:SDM_LGM}
\begin{align}
f_j(\boldsymbol{X})
&=
\boldsymbol{D}_j^{\mathsf T}(\boldsymbol{X}) \boldsymbol{u}_j,
\qquad
\boldsymbol{u}_j \mid \sigma^2_{A,j}
\sim
\mathcal{N}\!\left(
\boldsymbol{0},
\sigma^2_{A,j}\boldsymbol{Q}^{-}_{A,j}
\right),
\quad j = 1,\ldots,J
\label{eq:SDM_LGM_main}
\\
f_l(\boldsymbol{Z}, T)
&=
\boldsymbol{G}_l^{\mathsf T}(\boldsymbol{Z}, T) \boldsymbol{v}_l,
\qquad
\boldsymbol{v}_l \mid \sigma^2_{B,l}
\sim
\mathcal{N}\!\left(
\boldsymbol{0},
\sigma^2_{B,l}\boldsymbol{Q}^{-}_{B,l}
\right),
\quad l = 1,\ldots,L
\label{eq:SDM_LGM_spacetime}
\end{align}
\end{subequations}

Equations \eqref{eq:SDM_LGM_main}-\eqref{eq:SDM_LGM_spacetime} specify two sets of structured additive effects. Each abiotic component $f_j(\boldsymbol{X})$, $j=1,\ldots,J$, is represented through a basis $\boldsymbol{D}_j(\boldsymbol{X})$ and coefficients $\boldsymbol{u}_j$, which follow a zero-mean Gaussian prior with variance $\sigma^2_{A,j}$ and precision matrix $\boldsymbol{Q}_{A,j}$. Similarly, each biotic component is defined by basis $\boldsymbol{G}_l(\boldsymbol{Z},T)$, and coefficients $\boldsymbol{v}_l$ with variance $\sigma^2_{B,l}$ and precision matrix $\boldsymbol{Q}_{B,l}$. The variance parameters control the contribution of abiotic (A) and biotic (B) effects. In a Bayesian framework, the model specification is completed by a prior on the scale parameters $\sigma^2_{A,j},\sigma^2_{B,l}$: traditionally, the $\sigma^2_{A,j}$ parameters are fixed to large values to let the coefficients free to vary, while $\sigma^2_{B,l}$ are assigned independent vague priors to reflect lack of prior information.

Recently, \cite{F20} proposed an innovative method to perform prior specification of scale parameters under the name of Hierarchical Decomposition (HD) priors. This novel approach allows for a specification of a joint prior on the variances that can be easily built to reflect available prior knowledge about the relative importance of the model effects, as well as to regularize the flexibility of the model. The use of HD priors has been found to be competitive with respect to other state-of-the-art alternatives in \cite{hem2021robust} and \cite{marques2023variance}. HD priors are particularly valuable in fields characterized by structured data and/or the availability of expert knowledge. In settings with relatively standard designs (such as multilevel or spatio-temporal models), HD priors can be constructed to reflect the model structure and guided by the principle of parsimony, favoring simpler model solutions unless the data provide strong evidence for additional complexity. Alternatively, the HD prior design may also be guided by the principle of weak prior information, aiming to encode minimal assumptions beyond those implied by the model structure itself. Examples of the HD framework are illustrated in applications to disease mapping \citep{FV22,riebler2016intuitive}, demography (neonatal mortality in \cite{F20}), and forestry \citep{marques2023variance}. HD priors are also well suited to contexts where expert knowledge is naturally expressed in terms of variance proportions, including disease mapping \citep{wakefield2007disease}, genomics \citep{holand2013animal}, and ecology \citep{peres2006variation}, though direct elicitation has so far been applied only to plant breeding data \citep{hem2021robust}.

HD priors offer a practical solution to the issues concerning prior elicitations in Bayesian SDMs. First, a decomposition tree is immediately available due to the typical structure of these models where abiotic and biotic effects are modelled separately \citep{ovaskainen2017make}. Secondly, ecologists often have expert knowledge on variance partitioning \citep{pettit1990conditional}, which could be readily encoded in the model prior specification. Thirdly, constraining model flexibility is an important aspect and this can be more easily approached with principled priors, such as the HD ones.

This paper discusses the implementation of HD priors in the yet unexplored context of SDMs. Specifically, we outline the workflow required to implement the HD approach in SDMs, consisting of the steps of tree design and prior specification. Additionally, we emphasize the importance of the preliminary standardization process necessary for the proper interpretation of HD-based reparameterizations.

The proposal is exemplified through a real-world case study on the dataset provided by \cite{hui2023spatiotemporal}, which reports presence/absence data for 39 fish species in the North Atlantic from the NOAA-NEFSC survey \citep{NEFSC_FallBottomTrawlSurvey_2024}. The application demonstrates the advantages of the use of HD priors over more traditional approaches. Moreover, the complexity of the model required by this dataset provides an opportunity to highlight and discuss some of the critical challenges that may arise in practice during the specification of an HD prior. 


The rest of the paper is structured as follows: Section 2 introduces the motivating example, presenting the dataset and the chosen model; Section 3 briefly reviews the HD approach; Section 4 presents the proposed application of HD priors to SDMs; Section 5 reports the results from the case study; Section 6 concludes with a discussion.

\section{Motivating example}\label{sec:data}

We consider a dataset consisting of part of the NOAA-NEFSC fall bottom trawl survey database \citep{NEFSC_FallBottomTrawlSurvey_2024}, processed and studied in \cite{hui2023spatiotemporal} and made publicly available online at \url{https://github.com/fhui28/CBFM} (last access: January 2026). The main research question is concerned with the quantification of the variance contributions of the different model components, i.e. environmental factors and residual spatial and temporal contributions.

The dataset contains $N=5892$ observations collected over the U.S. Northeast continental shelf, during the fall season of each year from 2000 to 2019; each location has been visited only once. The occurrence of $N_{\text{species}}=39$ different demersal fish species is collected in the form of a presence/absence response $Y$. A visual overview of the data is reported in the Supplementary Material.

The following environmental covariates have been collected for each observation: \textit{Surface temperature} ($X_1$), \textit{Bottom temperature} ($X_2$), \textit{Surface salinity} ($X_3$), \textit{Bottom salinity} ($X_4$), \textit{Depth} ($X_5$). A binary covariate is also reported to represent the type of \textit{Survey vessel} ($X_6$) and is treated as an additional environmental condition, for a total of $P=6$ covariates. The spatial location is reported using the UTM coordinate system where $\boldsymbol{Z}=[Z_1,Z_2]$ are the latitude and longitude, while the $T$ covariate is set to represent the year, as the data has only been collected over the fall seasons.


In terms of modelling, each species' occurrence is treated marginally and fitted using an individual, although identical, SDM. 
The given SDM used here is inspired by the joint model proposed in \cite{hui2023spatiotemporal}, which assumed separability between the spatial effect, the temporal one and the effects of the continuous environmental covariates, based on exploratory analysis.
Dropping the indices for simplicity, each response $Y$ is assumed to follow a Bernoulli distribution with probability parameter $g^{-1}(\eta)$ where $g(\cdot)$ is the logit link and the linear predictor $\eta$ is defined as:
 \begin{gather}\label{eq:application_model}
\eta=\mu+\sum_{p=1}^5 f_{p}(X_p)+f_6(X_6)+f_S(\boldsymbol{Z})+f_T(T)
\end{gather}
The model is then specified as an LGM. Specifically, for the five continuous covariates $p=1,\dots,5$, $f_{p}(X_p)$ are modelled using P-Splines (\cite{LB}, \cite{EM}) in order to accommodate for possible smooth non-linear effects: in particular, the bases $	\mathbf{D}_p(X_p)$ are composed of 20 B-splines, while the precision matrices $\mathbf{Q}_{A,p}$ are specified as an intrinsic Gaussian Markov random field (IGMRF,  \cite{RH}) of order 2. $f_6(X_6)$ is a two-level i.i.d. effect on the dummy covariate; $f_T(T)$ is set to a first-order random walk (i.e. IGMRF of order 1). $f_S(\boldsymbol{Z})$ is specified using an ICAR model (i.e. an IGMRF prior of order 1) on a discretized grid underlying the spatial domain, whose coefficients are transformed to the continuous space using a two-dimensional B-spline basis to induce smoothness (more details in the Supplementary Material).
The model thus implicitly contains the following variance parameters: $\sigma^2_{1},\dots,\sigma^2_6,\sigma^2_S,\sigma^2_T$. 

\section{The Hierarchical Decomposition prior framework}
The HD prior framework \citep{F20} incorporates prior knowledge on the relative importance of model effects in an intuitive way via a two-step procedure. Consider the LGM presented in Equations \eqref{eq:SDM}-\eqref{eq:SDM_LGM} for clarity. The first step of the HD approach consists in considering a reparametrization of  $\sigma^2_{A,j},\sigma^2_{B,l}$ into a set of more intuitive parameters. Specifically, the new parameters include a single total variance:
\begin{equation}\label{eq:total_variance}
V=\sum_j \sigma^2_{A,j}+\sum_l \sigma^2_{B,l}
\end{equation}
and a set of proportions of variance denoted by $\omega$ letters.
The proportions are constructed using a tree design that recursively splits the total variance 
$V$ into groups of variance parameters, until each branch contains a single scale parameter. At each split, the proportions are computed by dividing the sum of scale parameters in the branches by the sum in their parent node. Binary splits yield scalar proportions, while multi-branch splits produce simplices. For example, in SDMs, users often have an intuition about the share of variance explained by abiotic effects. This can be represented by an initial binary split between abiotic and biotic effects, defining the proportion:
\begin{equation}\label{eq:omega_A}
\omega_{A}=\frac{\sum_j \sigma^2_{A,j}}{\sum_j \sigma^2_{A,j}+\sum_l \sigma^2_{B,l}}.
\end{equation}

The second step of the procedure consists in specifying prior distributions and their hyperparameters in a way that reflects prior beliefs on the new parameters $V$ and $\omega$s: the outcome is a joint prior distribution for the original variance parameters that effectively integrates expert knowledge, consistently with Bayesian principles. HD priors can be designed in a user-friendly way using the \texttt{makemyprior} R package developed by \cite{hem2021makemyprior}.


With respect to the proportion parameters, \cite{F20} consider two alternative prior specifications. When no prior preference exists between the branches of a split, Dirichlet priors are used to reflect ignorance about how the total variance is allocated. In contrast, particularly for binary splits, a user may express a prior preference for one branch, using a Penalized Complexity (PC) prior. The PC framework of \cite{S17} provides a principled approach to define priors that shrink towards a base model, i.e. a preferred parameter value. More precisely, a PC prior is defined by assigning an exponential distribution, with rate parameter $\lambda$, to a KLD-based distance between the model and its base version. Following \cite{hem2021makemyprior},  $PC_b(\cdot)$ denotes a PC prior where $b$ is the parameter value under the base model; for variance parameters, $PC_0$ is typically used to induce shrinkage towards the base model, i.e. having variance equal to 0. The rate $\lambda$ controls the degree of shrinkage and is calibrated through tail probability statements. PC priors are widely used, especially for Gaussian variance parameters like $V$ in Equation \eqref{eq:total_variance}, for which $PC_0$ reduces to an exponential prior on $\sqrt{V}$. 

In contrast, deriving a $PC_0$ prior for proportion parameters requires more effort, as its form depends on the specific effects involved and can also depend on lower-level proportion parameters, which is computationally inefficient \citep{F20}. Nevertheless, the $\text{PC}_0$ for proportions of variance $\omega$ for a latent Gaussian model is found in many instances \citep{F20,FV22} to have the following \textit{simplified form}:
\begin{gather}\label{eq:simplified_form}
\pi(\omega)=\frac{\lambda  \exp(-\lambda  \sqrt{\omega})}{2\sqrt{\omega}[1-\exp(-\lambda )]} \quad \quad 0<\omega<1
\end{gather}
which corresponds to an exponential distribution on $\sqrt{\omega}$, truncated at $\sqrt{\omega}=1$. This simplified solution does not depend on the choice of basis or precision matrices of effects, nor on the parameters at lower splits of the decomposition tree.  We prove in the Supplementary Material that Equation \eqref{eq:simplified_form} is the correct $PC_0$ prior for most practically relevant cases in SDMs. The proof we present is based on a condition on the involved matrices' ranks, whose validity depends on the type of effects considered.

The main advantage of the HD approach lies in transforming a variance-scale parametrization, which is difficult to interpret, into a proportion-scale one. This transformation enables the straightforward incorporation of prior information about the relative contributions of different effects through prior specification, making model assumptions more transparent and allowing ecologists and other non-expert users to reason directly in terms of intuitive effect importance rather than abstract variance components. However, the approach relies on the proportion parameters accurately reflecting the share of variance attributable to each effect within the branches relative to the total variance in the parent nodes. This condition holds only if the original effects have been correctly scaled so that their corresponding variance parameter matches their actual variance contribution: traditional specifications of effects do not always satisfy this requirement. 

To address this issue, \cite{ferrari2025} proposed a \emph{standardization procedure} applicable to all effects in an LGM. The standardization ensures that each effect's variance parameter (e.g. $\sigma^2_j$) accurately reflects the contribution to the total variance of the associated effect (e.g. $f_j(\cdot)$). This step is therefore essential for the interpretability of HD priors, as it guarantees that the proportion parameters derived from the decomposition tree meaningfully represent the relative importance of the effects. 

\section{HD priors for SDMs}

\subsection{Standardization of the SDM effects}\label{sec:stand_intro}
In order to correctly apply HD priors to an SDM model, it is necessary that all effects, abiotic and biotic, have been correctly standardized so that the parameters $\sigma^2_{A,j}$ and $\sigma^2_{B,j}$ correspond to the actual variance contributions of their respective effects. Based on \cite{ferrari2025}, this requirement can be formally defined as:
\begin{subequations}\label{eq:requirement}
\begin{align}
\sigma^2_{A,j} &= Var_{\boldsymbol{X},\boldsymbol{u}_j}\!\left[f_j(\boldsymbol{X}) \mid \sigma^2_{A,j}\right],
\qquad j=1,\dots,J, \label{eq:requirement_A} \\
\sigma^2_{B,l} &= Var_{\boldsymbol{Z},T,\boldsymbol{v}_l}\!\left[f_l(\boldsymbol{Z},T) \mid \sigma^2_{B,l}\right],
\qquad l=1,\dots,L. \label{eq:requirement_B}
\end{align}
\end{subequations}
As shown in \eqref{eq:requirement}, the variance contribution of an effect is defined in this framework as the variability of the effect with respect to the covariates involved. Consequently,  $\boldsymbol{X}$, $\boldsymbol{Z}$, and $T$ (other than the Gaussian coefficients $\mathbf{u}_j,\mathbf{v}_l$) must be treated as random variables with a known probability distribution for the quantities in \eqref{eq:requirement} to be formally defined. Importantly, the distributional assumptions made on the covariates  should not necessarily aim to reflect their true data-generating process; rather, they should be used to translate the user's understanding of the variance contributions of the effects into precise mathematical definitions.  Note that it is sufficient to specify the joint distribution of the covariates associated with a given model effect. The dependence structure between covariates that do not enter the same effect plays no role in the standardization procedure and may therefore be ignored; for convenience, sets of covariates that do not enter the same effect can be assumed to be independent. We suggest considering Uniformity as a default choice for the distributional assumption on the covariates and spatio-temporal domains. This recommendation is motivated by both mathematical convenience and interpretability. Specifically, Uniformity guarantees an interpretational advantage for the user: when a Uniform is chosen, the variance with respect to the covariate is simply the variance of the trend (i.e., realization of the effect) measured in all locations over the support, which is more intuitive for a user than having to take into account different probability densities at each point. 

The standardization procedure that guarantees \eqref{eq:requirement} modifies the basis and precision matrices of the effects through a two-step procedure, with both steps depending explicitly on the assumed distribution of the covariates. The first step consists of imposing appropriate constraints on the effects to ensure zero mean, whenever this is not already satisfied:
\begin{subequations}\label{eq:0mean}
\begin{align}
\mathbb{E}_{\boldsymbol{X}}\!\left[f_j(\boldsymbol{X})\right] &= 0,
\qquad j=1,\dots,J, \label{eq:0mean_A} \\
\mathbb{E}_{\boldsymbol{Z},T}\!\left[f_l(\boldsymbol{Z},T)\right] &= 0,
\qquad l=1,\dots,L. \label{eq:0mean_B}
\end{align}
\end{subequations}
This requirement can be enforced by suitably modifying either the precision matrix or the basis functions of the effects. 
The second step consists of scaling each effect by an appropriate constant, referred to as a reference standard deviation \citep{SR14}:
\begin{subequations}\label{eq:scaling}
\begin{align}
\widetilde{f}_j(\boldsymbol{X})
&= \frac{f_j(\boldsymbol{X})}
{\sqrt{\operatorname{Var}_{\boldsymbol{X},\boldsymbol{u}_j}
\!\left[f_j(\boldsymbol{X}) \mid \sigma^2_{A,j}=1\right]}},
\qquad j=1,\dots,J, \label{eq:scaling_A} \\
\widetilde{f}_l(\boldsymbol{Z},T)
&= \frac{f_l(\boldsymbol{Z},T)}
{\sqrt{\operatorname{Var}_{\boldsymbol{Z},T,\boldsymbol{v}_l}
\!\left[f_l(\boldsymbol{Z},T) \mid \sigma^2_{B,l}=1\right]}},
\qquad l=1,\dots,L. \label{eq:scaling_B}
\end{align}
\end{subequations}
Note that the variances in the denominators of Equations \eqref{eq:scaling} are computed conditioning respectively upon $\sigma^2_{A,j}=1$ and $\sigma^2_{B,l}=1$.
The scaling ensures that all effects $\widetilde{f}_j(\boldsymbol{X})$ and $\widetilde{f}_l(\boldsymbol{Z},T)
$ are expressed on a common scale. The scaling constants defined as reference standard deviations can be easily computed for many effects, including popular IGMRFs, using the \texttt{scaleGMRF} package available in GitHub: \url{https://github.com/LFerrariIt/scaleGMRF} \citep{ferrari2025}.
\subsubsection{Applying standardization to the NOAA-NEFSC case study}

Consider the model from Equation \eqref{eq:application_model}. In this case, the application of the standardization simply requires the choice of the marginal distributions $\prod_{p=1}^P\pi(x_p)\pi(\boldsymbol{z})\pi(t)$, since interaction terms are absent from the model. Starting from the environmental covariates, we choose to assume a Uniform distribution on each of them. In order to define the Uniform ranges for the continuous covariates, we consider the empirical ranges, after having excluded points deemed of high leverage (3 in total), which may greatly distort the results. The chosen ranges are coherent with the available knowledge about the behaviour of the considered environmental factors over the U.S. Northeast continental shelf \citep{NEFSC_FallBottomTrawlSurvey_2024}. Uniformity is also assumed on all values of the categorical variables $X_6$ and $T$. With respect to the spatial coordinates $\boldsymbol{Z}$, we also consider a Uniform on a polygon including all locations found in the sample and designed using a concave hull algorithm (see more details and figures in the Supplementary Material). 

Once the distributional assumptions have been specified, the standardization procedure can be carried out on all effects. However, dealing with P-Splines poses an additional challenge, as the variance parameter of a second-order IGMRF (employed in the traditional specification of P-Splines) only measures the deviation of the effect from its null space, which is its linear trend. Hence, in this case, the $\sigma^2_1,\dots,\sigma^2_5$ parameters only measure the deviation of the effects from their linear trends, rather than quantifying the overall contribution to the variance of their corresponding effect. As a solution, \cite{ferrari2025} propose to respecify the P-Spline effects with two separate components: a linear effect, and a P-Spline effect constrained so as not to interfere with the linear one. Thus, each $f_p(X_p)$ is redefined as:
\begin{equation}
f_p(X_p)=f_{Lp}(X_p)+f_{Np}(X_p)
\end{equation}
where $f_{Lp}(X_p)$ is a linear effect with corresponding variance parameter $\sigma^2_{Lp} $ and $f_{Np}(X_p)$ is a P-Spline effect (under appropriate constraints removing the linear trend) with corresponding parameter $\sigma^2_{Np} $. See \cite{ferrari2025} for more details on this alternative specification.

Finally, standardization has been separately applied to all $f_{Lp}(X_p)$ and  $f_{Np}(X_p)$ for $p=1,\dots,5$, as well as to $f_6(X_6),f_{\boldsymbol{Z}}(\boldsymbol{Z}),f_T(T)$.

\subsection{Tree design}\label{sec:tree_design}

\begin{figure}[p]
    \centering
\includegraphics[width=\textwidth]{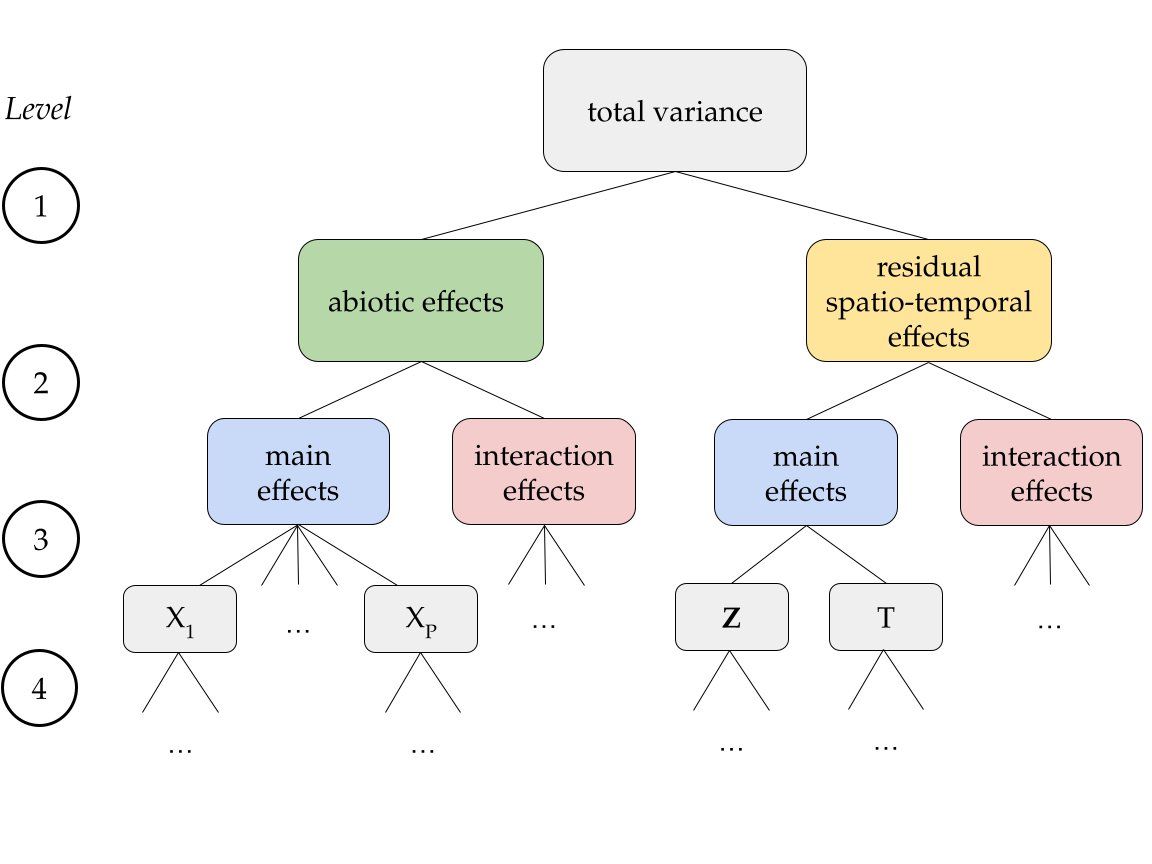}
    \caption{Default decomposition tree for SDMs. Each node represents the sum of all the variance parameters $\sigma^2$ from the corresponding effects.}
    \label{fig:default_tree}
\end{figure}

After appropriate standardization of all effects in an SDM, we can consider the application of the HD framework, starting from the tree design. According to \cite{F20}, the \textit{``tree structure must be selected so that the desired comparisons can be made''}: this means that the resulting proportion parameters should measure the relative importance between groups of effects for which the user has a direct intuition. The tree design should therefore be ideally application-specific and expert-driven. However, it is often the case that the available prior information is not directly pertinent to the case study at hand, but rather comes from the combination of general field knowledge and adoption of modelling principles (e.g. simplicity). For this reason, we propose a default decomposition tree for the generic SDM of Equation \eqref{eq:SDM_LGM} represented in Figure \ref{fig:default_tree}, designed to offer parameters for which ecologists have a broad intuition, regardless of the application at hand. This default tree design is based on the theory behind SDMs, modelling principles, and considerations from the HD literature. 

The first split in the tree from Figure \ref{fig:default_tree} partitions the total variance into the two main sources of variability recognized by the SDM theory: the contribution of abiotic/environmental effects; the contribution of the additional spatio-temporal variability that could represent biotic interactions and stochastic processes. At the second level, there is a split for each of the two child nodes from the previous step. In both cases, the main effects (functions of single covariates) are separated from interaction terms (functions of two covariates); note that $\boldsymbol{Z}$ is considered a single covariate. If no interactions are included in the model, this level can simply be pruned from the tree. The third level involves potential multi-branch splits, dividing effects on the basis of the covariates they are functions of. A first split is used to divide the effects of the different environmental covariates into multiple branches, while a second split separates the spatial effects from the temporal one. Interaction terms from either branch of Level 2 might require additional splits at Level 3.

The proposed tree well adapts to most SDM applications. The tree can then be tweaked at will, whenever the user possesses more sophisticated information. If all the grey nodes in Figure \ref{fig:default_tree} contain a single variance parameter, then the tree is complete as it is. However, some covariates may enter the model through multiple effects. In such cases, the scheme in Figure \ref{fig:default_tree} should be extended with additional splits until all child nodes are singletons. We recommend using fourth-level splits that separate effects by increasing flexibility via successive binary branches, as this simplifies prior specification. For example, a Level 4 split may distinguish between the linear and non-linear components of an environmental covariate effect.

\subsubsection{Tree design for the NOAA-NEFSC case study}
\begin{figure}[p]
    \centering
    \includegraphics[width=\textwidth]{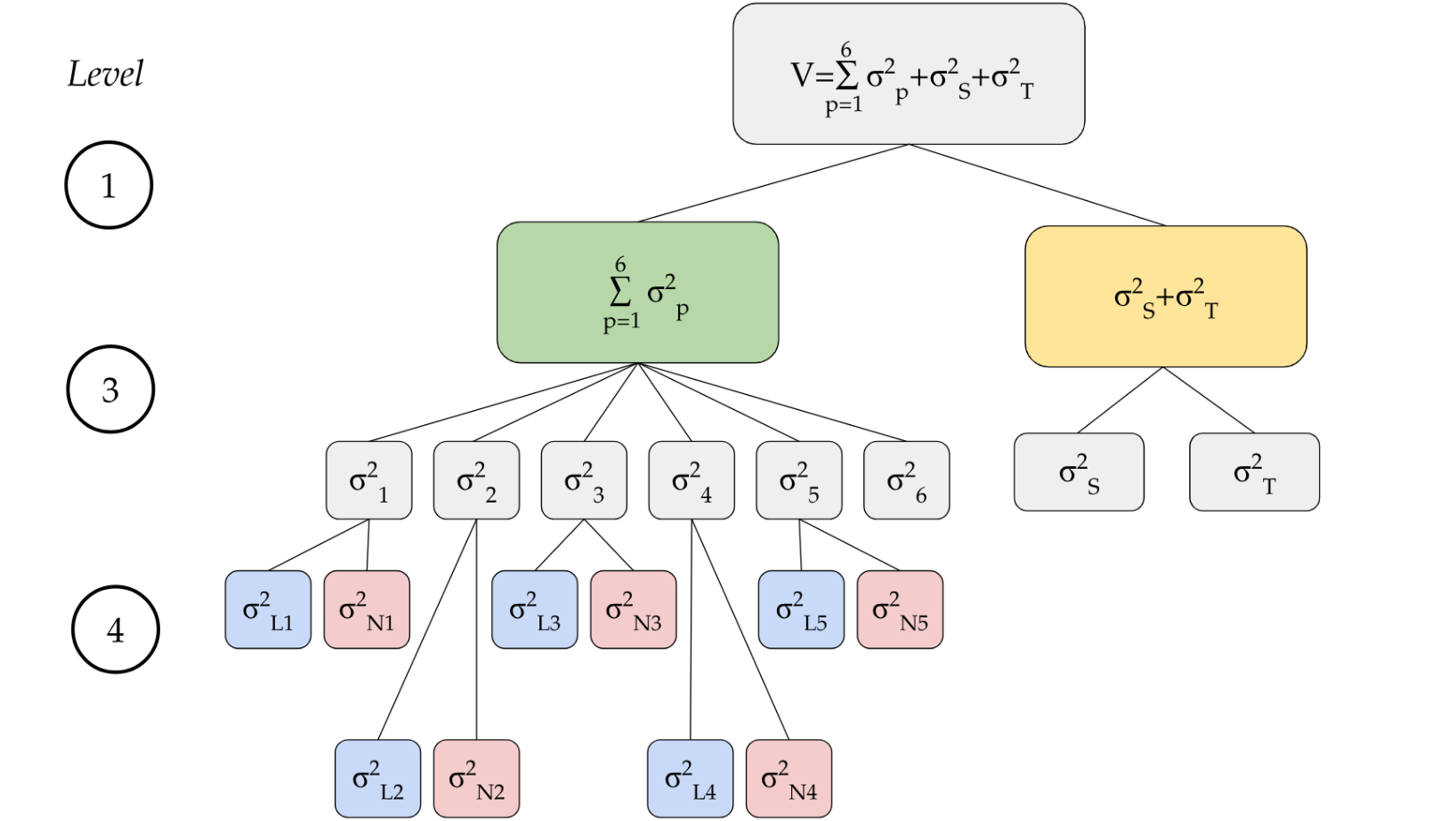}
    \caption{Decomposition tree for the case study based on the default proposal of Section \ref{sec:tree_design}.}
    \label{fig:application_tree}
\end{figure}

The model for the case study has a total of 13 variance parameters that need prior specification: $\sigma^2_{L1},\sigma^2_{N1},\dots,\sigma^2_{L5},\sigma^2_{N5},\sigma^2_{6},\sigma^2_{S},\sigma^2_{T}$. The tree design for this application is represented in Figure \ref{fig:application_tree}. Level 2 of the default tree is pruned, while, at Level 4, a binary split between the linear and non-linear contribution is added for all the 5 environmental covariates that are continuous. 
We obtain the following new parameters dividing the child nodes by their parent node:

\begin{subequations}\label{eq:cs_new_pars}
\begin{align}
V
&= \left(\sum_{p=1}^5 \sigma^2_{Lp}+\sigma^2_{Np}\right)
   + \sigma^2_6 + \sigma^2_S + \sigma^2_T
   \label{eq:cs_new_pars_V} \\
\omega_{A}
&= \dfrac{\left(\sum_{p=1}^5 \sigma^2_{Lp}+\sigma^2_{Np}\right)
   + \sigma^2_6}{V}
   \label{eq:cs_new_pars_omegaA} \\
\boldsymbol{\omega}_X
&= \left[
\dfrac{\sigma^2_{L1}+\sigma^2_{N1}}{\omega_A \cdot V},
\ldots,
\dfrac{\sigma^2_{L5}+\sigma^2_{N5}}{\omega_A \cdot V},
\dfrac{\sigma^2_6}{\omega_A \cdot V}
\right]
\label{eq:cs_new_pars_omegaX} \\
\omega_S
&= \dfrac{\sigma^2_S}{\sigma^2_S+\sigma^2_T}
\label{eq:cs_new_pars_omegaS} \\
\omega_{Np}
&= \dfrac{\sigma^2_{Np}}{\sigma^2_{Lp}+\sigma^2_{Np}},
\qquad p=1,\ldots,5.
\label{eq:cs_new_pars_omegaNp}
\end{align}
\end{subequations}

\subsection{Prior specification}

Based on Figure \ref{fig:default_tree}, we discuss suitable prior choices for the new HD parameters, i.e. $V$ and $\omega$s. Although the original HD framework constructs joint priors in a bottom-up fashion, most practical implementations assume independence between proportions across splits. We therefore adopt prior independence between the split-specific parameters, as this choice substantially simplifies prior specification and computation while remaining consistent with common practice.

\paragraph{Total variance}
With respect to $V$, PC priors are recommended in the HD literature when the likelihood is non-Gaussian; for Gaussian likelihoods, \cite{F20} instead advocates for the scale-invariant Jeffreys prior. In the context of SDMs, both choices are appealing, as $\text{PC}_0$ priors offer regularization for highly flexible models, while Jeffreys priors avoid the need for hyperparameter tuning. 

\paragraph{Level 1} 
Level 1 produces the single proportion parameter $\omega_{A}$ from Equation \eqref{eq:omega_A}. Ignorance can be easily expressed through a Uniform distribution or a scale-invariant Jeffreys prior. If prior information suggests otherwise, this can be introduced using an appropriate Beta distribution, or a PC prior with the desired base model and concentration parameter reflecting the user's uncertainty \cite{hem2021makemyprior}. 

\paragraph{Level 3}
The splits of Level 3 divide the effects from the parent nodes into branches according to the covariates they represent. 
\cite{F20} suggest the use of a symmetric Dirichlet $\text{Dir}(q,\dots,q)$, where the hyperparameter $q$ controls the sparsity level and can be regulated to reflect assumptions about the partition of the variance among the branches. 
\cite{hem2021makemyprior} recommend to set $q$ using the marginal prior on a single proportion $\omega_p$ such that $P(\text{logit}(1/4)<\text{logit}(\omega_p)-\text{logit}(1/P)< \text{logit}(3/4))=0.5$. Alternatively, the work of \cite{R2D2} derived a desirable shrinkage behaviour on the induced prior on the coefficients of the R2D2 prior under the choice of $q<1/2$.

\paragraph{Levels 2 \& 4}
Levels 2 and 4 induce analogous binary splits that separate effects according to their degree of flexibility (e.g., main versus interaction effects, or linear versus non-linear effects). Following the guidelines of \cite{F20}, a natural choice is to interpret the proportion parameter as the proportional contribution of the more flexible branch and to assign it a $\mathrm{PC}_0$ prior, in line with the principle of model simplicity. In most cases, the resulting $\mathrm{PC}_0$ prior reduces to the simplified form given in Equation~\eqref{eq:simplified_form}. In situations where this simplification does not hold, we nevertheless advocate using the same functional form for practical reasons, as it offers substantial computational advantages, even though it no longer corresponds to the exact PC prior.

\subsubsection{Prior specification for the NOAA-NEFSC case study}
The parameter $\omega_A$ represents the proportion of variance explained by abiotic factors relative to biotic ones. The parameter $\omega_S$ quantifies the relative contribution of spatial versus temporal effects. Given the unavailability of prior information on these parameters, we assign vague Uniform priors to both parameters. This choice reflects no a priori preference for any admissible value by treating all variance partitions as equally plausible, it implies an expectation of equal partition while accommodating for substantial uncertainty.
On the other hand, $\boldsymbol{\omega}_X$ weighs the proportional contributions of each of the 6 environmental covariates, quantities for which it was possible to formulate prior assumptions. Specifically, \textit{Depth} was believed to play an important role, while the correlation between \textit{Surface temperature} and \textit{Bottom temperature}, as well as between \textit{Surface salinity} and \textit{Bottom salinity} may cause only one of the two covariates in each pair to be selected; finally, we do not have prior knowledge about the potential role of the \textit{Survey Vessel}. We chose to quantify these prior beliefs by setting $\boldsymbol{\omega}_X\sim \text{Dir}(q)$ with $q=0.5$, which induces moderate shrinkage in the covariate effects.

The 5 splits at Level 4 all separate the linear and non-linear contribution of the 5 environmental covariates. For each of the $\omega_{Np}$ parameters, we choose a highly flat $\text{PC}_0$ prior to induce little shrinkage of the non-linear contribution a priori. From the proof in the Supplementary Material, we know that the $\text{PC}_0$ prior has its simplified functional form in this case, i.e., Equation \eqref{eq:simplified_form}. The simplification is guaranteed here by using a reduced-rank representation for the non-linear effect, specifically with 20 B-spline basis functions.
To set the hyperparameter, we consider a tail probability statement: $P(\omega<U)=\alpha$. \cite{FV22} noted that the following condition must be respected: $\alpha\geq \sqrt{U}$ to obtain $\lambda>0$. As a consequence, the median can be at most $0.25$, which can be obtained with $\lambda \rightarrow 0$. We choose $\lambda=0.1$, which results in the median ($\alpha=0.5$) to be $0.238$, reflecting that a priori only about 1/4 of the variance contributed by the covariate effect comes from its non-linear component. The prior specification is completed assuming a scale-invariant Jeffreys prior on $V$.

\section{Results}
\subsection{Performance evaluation}
The performance of the proposed HD prior is first compared to more traditional choices, specifically: independent Inverse-Gamma(1,5e-5) on all the $\sigma^2$ parameters, which is the default specification in INLA (IG); independent $\text{PC}_0(\lambda)$ prior on all the $\sigma^2$ parameters with hyperparameter $\lambda$ such that $P(\sigma>3)=0.05$ as in \cite{F20} (PC).

\begin{figure}[p]
    \centering
    \includegraphics[width=\textwidth]{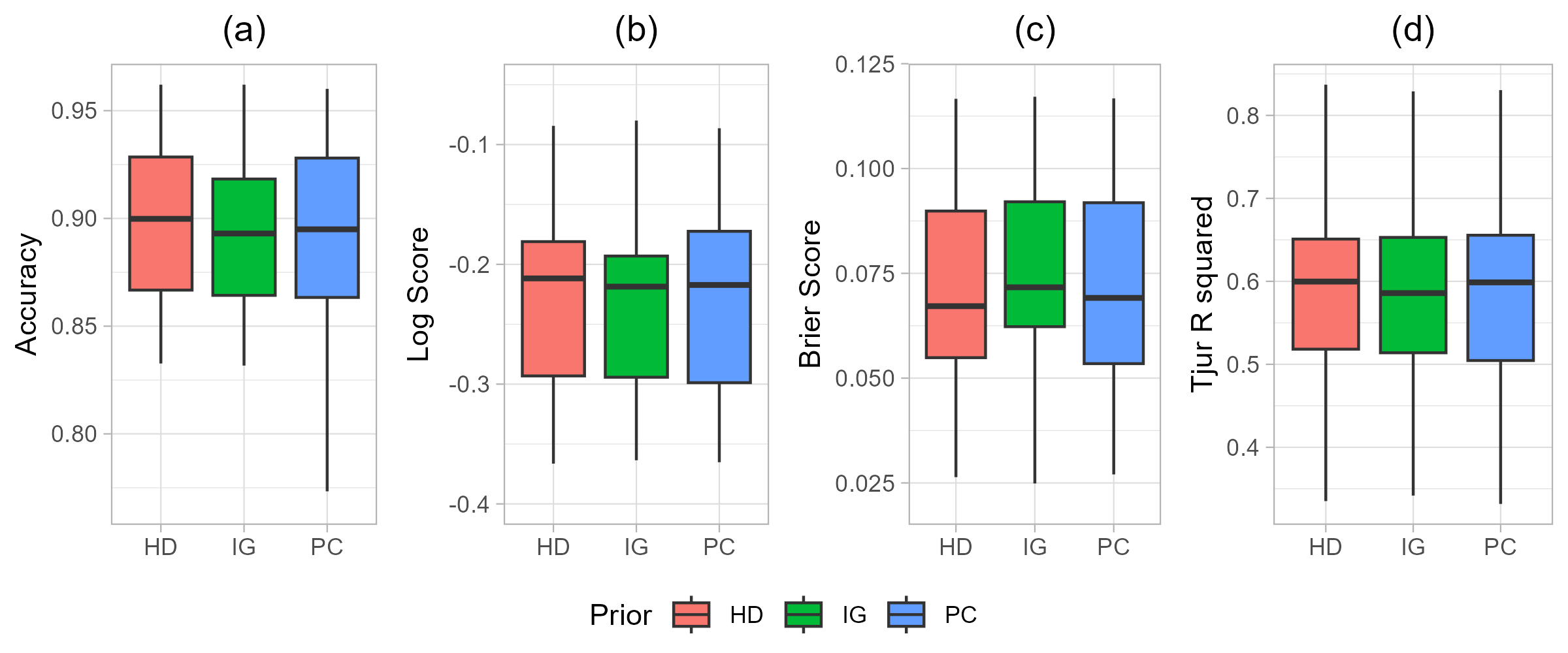}
    \caption{Comparison of the performance of the 4 different prior specifications on prediction on the test set.}
\label{fig:performance_metrics}
\end{figure}

The models are fitted in INLA \citep{rue2009inla}, which offers a joint posterior sample of all the model parameters, along with many other useful outputs. The dataset is divided into $\boldsymbol{y}=[\boldsymbol{y}_{\text{train}}^T,\boldsymbol{y}_{\text{test}}^T]^T$ where the training set $\boldsymbol{y}_{\text{train}}$ contains all observations up to 2015, while the test set $\boldsymbol{y}_{\text{test}}$ contains the remaining ones up to 2020, for a total of 1028 observations or approximately $17\%$ of the total dataset. 
The models are fitted on the training sets and the performance in terms of prediction over the test set is evaluated using the same criteria used in \cite{hui2023spatiotemporal} (log likelihood, Brier score, Tjur $R^2$), along with the more interpretable accuracy metric.
All metrics are based on the point estimates $\hat{p}_i=\text{logistic}(E[\eta_i|\boldsymbol{y}_{\text{train}},\boldsymbol{x}_i,\boldsymbol{z}_i,t_i])$ for all instances $i$ in the test set. 
Figure \ref{fig:performance_metrics} reports the distribution of these 4 metrics across all the 39 fish species in the dataset.

At first, we can note how the predictive performance is not largely impacted by the prior specification, thanks to the large number of observations in the dataset. Nevertheless, there are still some differences in the results: the HD prior is competitive with the popular PC prior choice, and represents an improvement to the Inverse-Gamma, particularly noticeable in terms of accuracy and Brier score. 
Hence, these results suggest that HD priors provide a credible alternative to prior specifications previously validated in the literature, at least within the context of this example.

\subsection{Variance partitioning results}
Using the HD prior, we can now estimate the variance contributions of the different effects on all the species' occurrence, which is a core research question for ecologists. We fit the models to the full dataset for each species and then a posterior sample on the model parameters is obtained using the \texttt{inla.posterior.sample()} function of the R-INLA package \citep{rue2009inla}. We estimate the variance contributions of each effect using finite-population variances computed with respect to the the Uniform distribution chosen on the covariates: $s^2_{Lp}=Var_{X_p}[f_{Lp}(X_p)|\beta_p], p=1,\dots,5;
s^2_{Np}=Var_{X_p}[f_{Np}(X_p)|\boldsymbol{u}_p], p=1,\dots,5;
s^2_{6}=Var_{X_6}[f_6(X_6)|\beta_6];
s^2_{S}=Var_{\boldsymbol{Z}}[f_S(\boldsymbol{Z})|\boldsymbol{v}_S];
s^2_{T}=Var_{T}[f_T(T)|\boldsymbol{v}_T]$.

\begin{figure}[p]
    \centering
    \includegraphics[width=\textwidth]{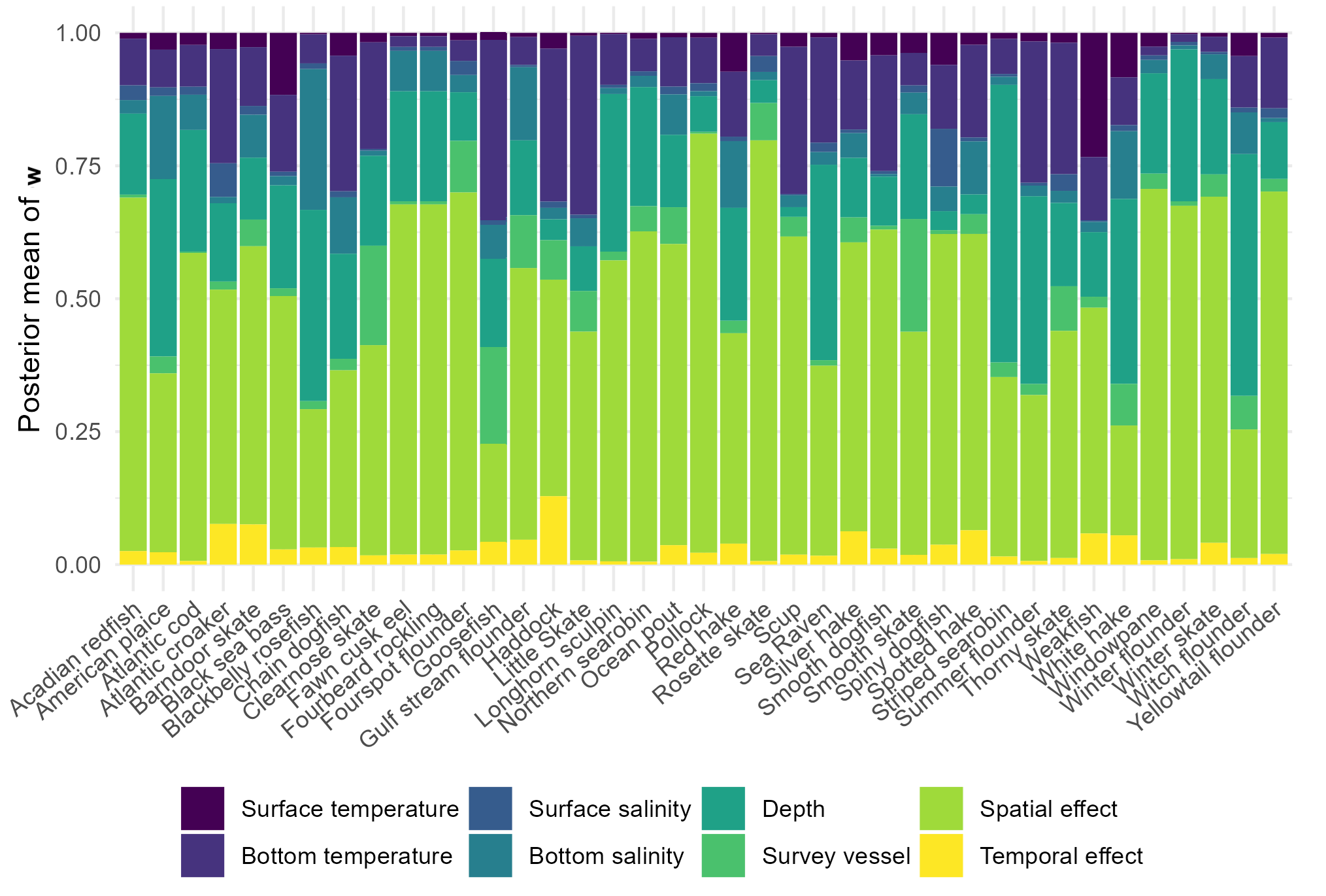}
    \caption{Posterior means of the entries of $\boldsymbol{\phi}$ for the 39 different species.}   \label{fig:application_vp}
\end{figure}

Letting $s^2_{p}=s^2_{Lp}+s^2_{Np},\;p=1,\dots,5$, we then compute the proportions of the total variance for each model terms:
\begin{gather}
\boldsymbol{\phi}=\dfrac{1}{\sum_{p=1}^6s^2_{p}+
s^2_S+
s^2_T}\left[s^2_1,\dots,s^2_6,
s^2_S,
s^2_T\right].
\end{gather}
Figure \ref{fig:application_vp} reports the posterior mean of the entries of $\boldsymbol{\phi}$ for all 39 species. 

Our analysis indicates that the primary factors influencing occurrence variability are the spatial effect, the \textit{Depth} effect, and the \textit{Bottom temperature} effect. This is consistent with previous studies \citep{hui2023spatiotemporal,kleisner2016effects}, as the selected species are all demersal and depth is a well-known driver of habitat suitability, often acting as a proxy for multiple environmental gradients such as pressure, light availability, and substrate type. Bottom temperature is also a key determinant of demersal species distributions, reflecting physiological constraints and metabolic preferences. 

\subsection{Prior sensitivity analysis}
We finally consider how the HD approach makes the process of prior sensitivity analysis more transparent and intuitive compared to traditional prior specifications. In standard approaches, the impact of hyperparameter choices on model behavior is often indirect and difficult to interpret, making it challenging for users to translate substantive prior beliefs into concrete modeling decisions. 

\begin{figure}[p]
    \centering
\includegraphics[width=\textwidth]{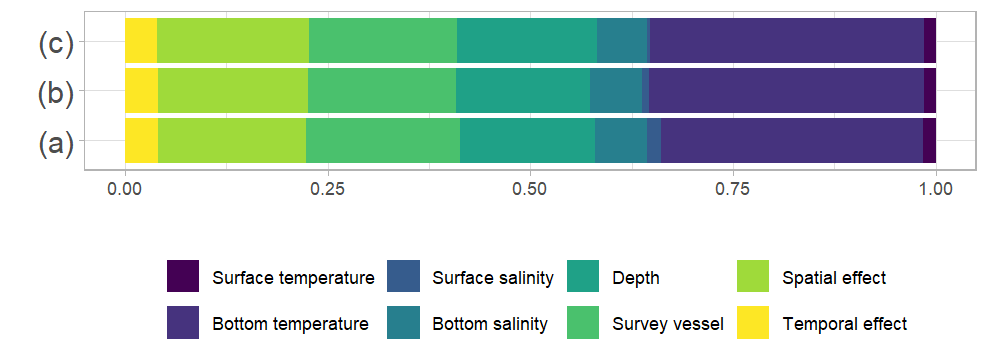}
    \caption{Posterior means of the entries of $\boldsymbol{\phi}$ for \textit{Goosefish}: (a) $q=1$, (b) $q=0.5$, (c) $q=1/6$.}
    \label{fig:prior_sensitivity}
\end{figure}

\begin{figure}[p]
    \centering
\includegraphics[width=\textwidth]{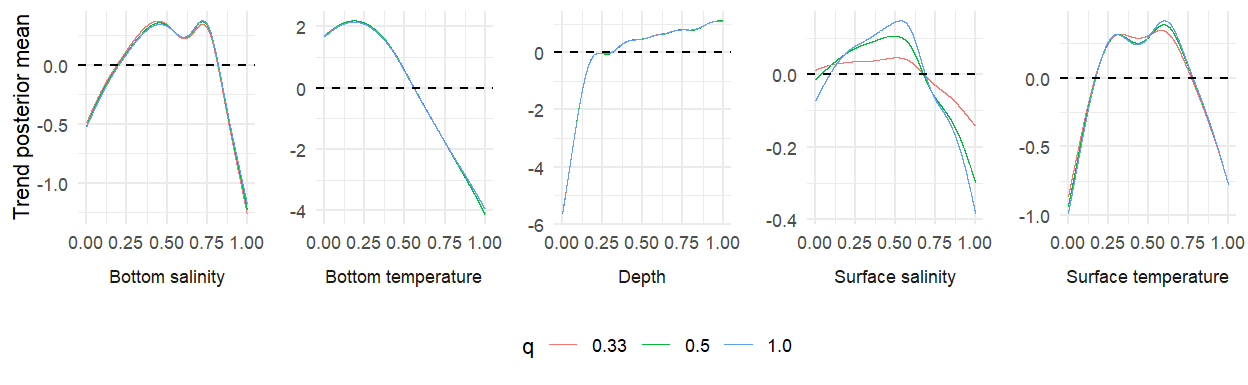}
    \caption{Posterior mean of $f_p(X_p),\; p=1,\dots,5$ for the 5 environmental covariates (in a normalized scale) and different values of $q$.}
    \label{fig:sensitivity_trends}
\end{figure}

In contrast, the HD framework allows practitioners to directly control specific aspects of the prior, such as the degree of shrinkage imposed on variance components. This interpretability facilitates prior sensitivity analysis, enabling a more direct assessment of the model's robustness/sensitivity to particular prior assumptions. Here, we illustrate this point by fitting the same model using the same HD prior structure while varying only the hyperparameter $q$ of the prior on $\boldsymbol{\omega}_X$, which controls the amount of shrinkage applied to the proportional variance contributions of the environmental covariates. Figure \ref{fig:prior_sensitivity} presents the resulting variance partition estimates for $q=[1,0.5,1/6]$ for the species \textit{Goosefish} (one of the most occurrent species). As $q$ decreases, the prior increasingly favors sparser variance allocations by progressively shrinking smaller components towards zero. This is confirmed when we look at the estimated trends for the different environmental covariates in Figure \ref{fig:sensitivity_trends}, where the trend of \textit{Surface salinity}  is shrunk closer and closer to 0. The observed behavior is both predictable and interpretable, highlighting how the HD approach offers more intuitive control over the prior influence, thereby facilitating principled prior sensitivity analyses and improving transparency in Bayesian modeling.

\section{Discussion}

In this paper, we introduced the HD prior framework to the context of SDMs, with the goal of enhancing the interpretability and transparency of prior specification through an explicit variance-partitioning perspective. By reparametrizing variance components into a total variance and sets of proportions, the framework provides a principled way to express prior beliefs about the relative contributions of environmental covariates and spatial and temporal effects in SDMs. We first proposed a default HD tree, designed by combining SDM theory, general modelling principles, and insights from previous HD applications. This default structure is sufficiently general to accommodate a broad range of SDM settings, while remaining flexible enough to incorporate expert knowledge when available.  We also provided practical guidelines for prior specification on the resulting reparametrization. In this respect, the availability of a simplified closed-form Penalized Complexity prior for variance proportions substantially reduces implementation complexity. Looking ahead, this default strategy could be extended to other application domains with similar model structures, such as disease mapping and environmental quality assessment.

The application to the NOAA-NEFSC dataset allowed us to demonstrate the use of HD priors in a complex modelling setting. We first showed how standardization can be applied to each model component to obtain a coherent interpretation of the associated variance parameters. We then highlighted the flexibility of the default tree, illustrating how it can be pruned or expanded to suit the specific features of a given study. At the prior specification stage, we proposed the use of $\text{PC}_0$ priors for the proportion of variance attributed to the non-linear component of P-spline effects. This choice merits further investigation, particularly to assess its potential as a general mechanism for encouraging model simplicity while preserving sufficient flexibility. In terms of predictive performance, HD priors performed comparably to commonly used alternatives. Importantly, the HD framework enables transparent and targeted prior sensitivity analysis, an aspect lacking in conventional specifications. Overall, HD priors represent a valuable addition to the SDM literature, offering substantially greater control and clarity in how variance is allocated a priori across environmental, spatial, and temporal effects.

An additional interesting aspect of this work concerns the proposed alternative way to quantify variance contributions. Rather than relying on the traditional approach based on the empirical distribution of the covariates \citep{ovaskainen2020joint}, we consider using user-specified distributional assumptions on them for the quantification of posterior quantities. The empirical distribution can be seen as a special case within this framework, but it lacks robustness and does not allow meaningful comparisons across case studies. Conversely, employing sample-independent choices, such as uniform distributions, improves the interpretability of variance contributions for the users and is a portable approach to different case studies. Further exploration of this alternative formulation could be an interesting direction for future work.

Finally, we note that the stacked modelling approach, which we adopted for the analysis of the case study, is restrictive and does not reflect contemporary community ecology theory, which emphasizes the role of biotic interactions in species occurrence. Current practice in this field increasingly relies on joint species distribution models (JSDMs; \citep{warton2015so}, which explicitly account for residual correlations among species. An important direction for future research is therefore to investigate the applicability of HD priors in the context of JSDMs. 

\paragraph{Data Availability}
Data supporting the findings of this study are publicly available in the GitHub repository at \url{https://github.com/fhui28/CBFM}.

\paragraph{Funding Information} The authors were supported by the European Union under the NextGeneration EU Programme within the Plan “PNRR - Missione 4 “Istruzione e Ricerca” - Componente C2 Investimento 1.1 “Fondo per il Programma Nazionale di Ricerca e Progetti di Rilevante Interesse Nazionale (PRIN)” by the Italian Ministry of University and Research (MUR), Project title: “METAbarcoding for METAcommunities: towards a genetic approach to community ecology (META2) ”, Project code: 2022PA3BS2 (CUP E53D23007580006), MUR D.D. financing decree n. 1015 of 07/07/2023.

\bibliographystyle{apalike}
\bibliography{sn-bibliography}

\newpage

\begin{center}
\Large\bfseries
Supplementary Material \\[0.3em]
for \textit{Bayesian Species Distribution Models Using Hierarchical Decomposition Priors} \\[0.5em]
\end{center}

\setcounter{section}{0}
\renewcommand{\thesection}{S\arabic{section}}

\section{Spatial effect specification}
\begin{figure}[p]
   \centering
\includegraphics[width=\textwidth]{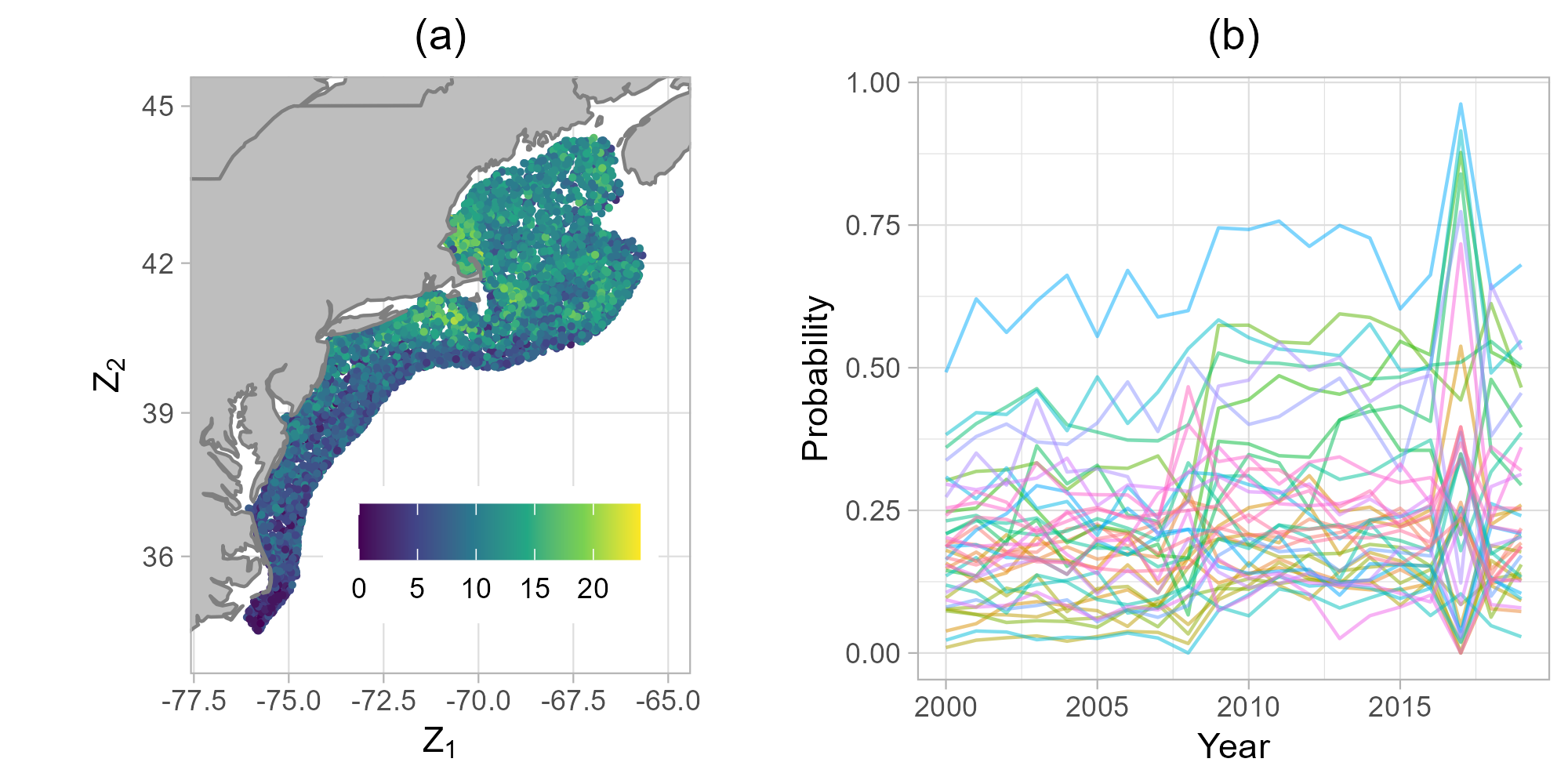}
    \caption{Overview of the case study dataset: (a) number of different species detected in each location; (b) proportions of presence observations for each of the 39 species over time.}
    \label{fig:case_study_eda}
\end{figure}

Consider the NOAA-NEFSC dataset presented in Section \ref{sec:data}. Figure \ref{fig:case_study_eda} offers a spatio-temporal overview of the data.

\begin{figure}[p]
    \centering
\includegraphics[width=\textwidth]{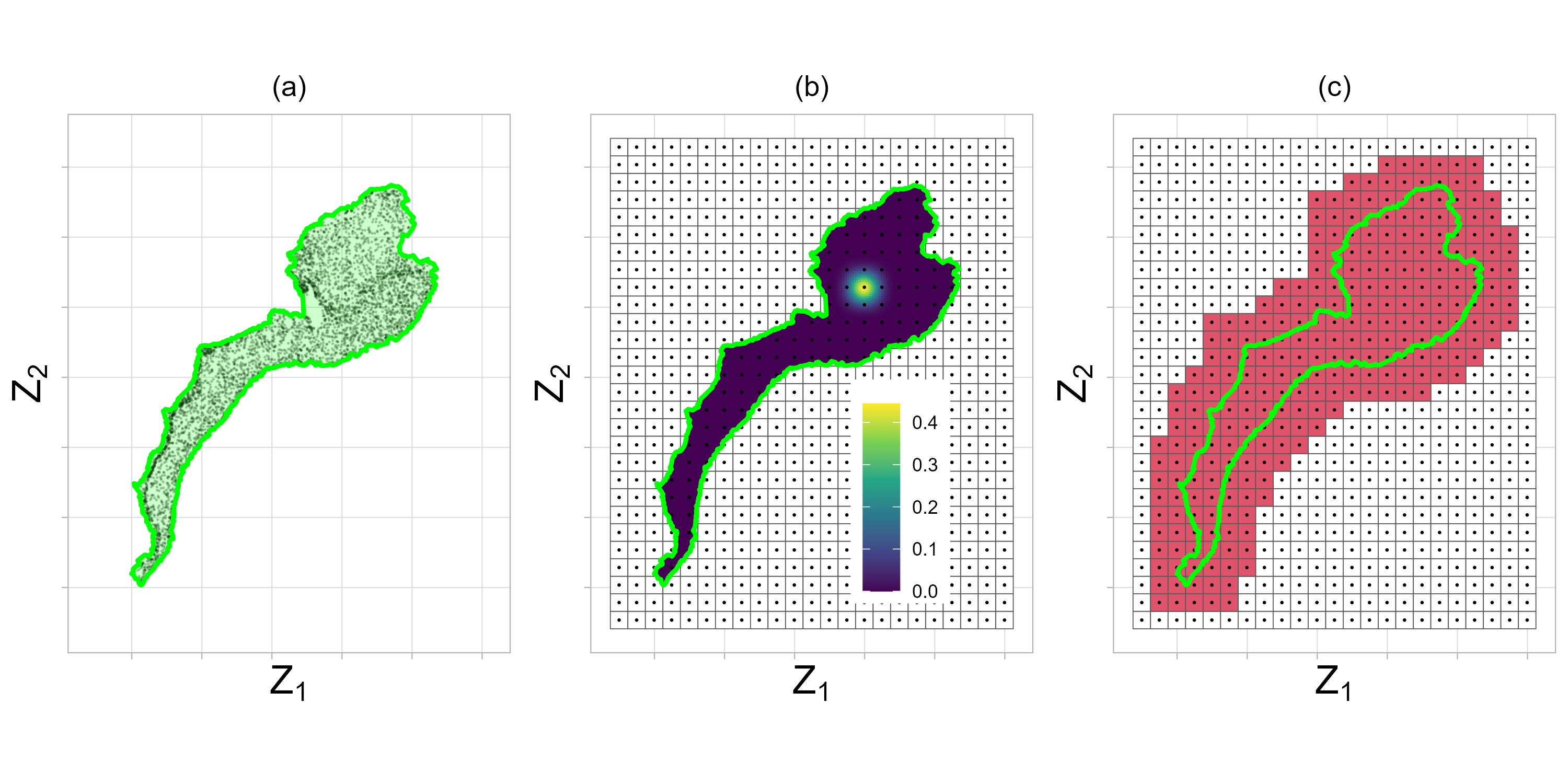}
    \caption{(a) Representation of the sampling locations over the $Z_1,Z_2$ spatial coordinates system with black dots, along with the chosen concave polygon containing all data points, outlined and shaded in green. (b) 50x50 km grid with cell centroids (black dots), along with the value of one B-Spline basis function centered in a grid cell over the area of interest. (c) The red grid cells indicate that the corresponding basis functions (centered in their centroids) have non-null values over the polygon represented by the solid green line.}.
    \label{fig:spatial_maps}
\end{figure}

The spatial locations in the sample all belong to a specific area of the North Atlantic ocean called U.S. North-east continental shelf. As it can be seen from Figure \ref{fig:spatial_maps} (a), the area has not been uniformly sampled, with some regions more thoroughly sampled than others. We consider a Uniform distribution which equally distributes importance all over an area of interest, and thus allows for a more intuitive definition of variance contribution attributable to the spatial component. Specifically, we define the \textit{area of interest} as a polygon that includes all locations found in the sample. The polygon is designed to cover all the observed spatial locations and found using a concave hull algorithm \citep{concaveman}: the resulting shape is represented in Figure \ref{fig:spatial_maps} (a) as the shaded green area. To quickly approximate desired quantities with respect to this distribution, a large sample of points equally distributed over the polygon area is then generated.

Once the distribution for $\boldsymbol{Z}$ has been well defined, the spatial effect can be then specified and scaled accordingly. Here, we choose two-dimensional P-splines (\cite{LB}) to model the variability over the continuous spatial coordinates. In the dataset, the distance between locations highly varies so that multiple resolutions could be used to model the spatial trend. Here we choose to capture a fairly large-scale spatial pattern creating a grid of 50x50 km cells and centering a 2D B-Spline in each cell. A two-dimensional B-Spline basis can be simply created using the Kronecker product between two univariate B-Spline basis (in this case both with 50 km distance between basis functions). Figure \ref{fig:spatial_maps} (b) shows the grid and the values of one of the basis functions over the polygon of interest. With respect to the precision matrix, we could simply use an IGMRF of first-order derived considering the overall rectangular grid. However, this method would be appropriate only if we actually want to model the whole rectangular surface. In this case, however, the interesting area is only the irregular polygon. Thus, we decide to remove all basis functions that have null values all over the polygon area. The grid cells whose corresponding basis functions have been retained ($K_S=267$) are colored in red in Figure \ref{fig:spatial_maps} (c). From this red grid or lattice, we can retrieve an adjacency matrix $\boldsymbol{W}$ that reports the neighbours of each remaining basis function. $\boldsymbol{W}$ is used to create a precision matrix $\boldsymbol{Q}_S$, which corresponds to an IGMRF of first order over the irregular lattice highlighted in Figure \ref{fig:spatial_maps} (c):
\begin{gather*}
\boldsymbol{Q}_S=\text{diag}(\boldsymbol{W1})-\boldsymbol{W}.
\end{gather*}
The model can therefore be defined as:
\begin{gather*}
f_S(\boldsymbol{Z})=\boldsymbol{B}^T(\boldsymbol{Z})\boldsymbol{v}_S,\\
\boldsymbol{v}_S|\sigma^2_S \sim N\left(\boldsymbol{0},\sigma^2_S \dfrac{\boldsymbol{Q}^*_S}{C_S}\right) 
\end{gather*}

$\text{ subject to } \boldsymbol{S}_S^T\boldsymbol{v}_S=\boldsymbol{0}$
where $\boldsymbol{B}(\cdot,\cdot)$ is the bivariate B-Spline created as detailed above, and $\boldsymbol{S}_S$ is the null space of $\boldsymbol{Q}_S$. 

However, the constraint $\boldsymbol{S}^T_S \boldsymbol{v}_S =0$ does not imply $E_{\boldsymbol{Z}}[f_S(\boldsymbol{Z})|\boldsymbol{v}_S]=0$ but only $\sum_{k=1}^{K_S} v_{S,k}=0$.
This means that the effect is subject to a constraint that is not directly interpretable in the scale of the spatial coordinates, which is inconvenient and creates an identifiability issue with the intercept parameter $\mu$. Following \cite{ferrari2025}, we replace the precision matrix with a modified version $\widetilde{\boldsymbol{Q}}_S$ with null space $\widetilde{\boldsymbol{S}}_S$ such that $\widetilde{\boldsymbol{S}}_S^T\boldsymbol{v}_S=0\implies E_{\boldsymbol{Z}}[f_S(\boldsymbol{Z})]=0$, which is more convenient.

\section{Proof of simplified form of the $PC_0$ prior for proportions}
Let $X_0$ and $X_1$ be two generic covariates, which can also be multivariate and can also represent spatio-temporal covariates. Let $N_0$ be the cardinality of the set of values $x\in\mathcal{X}_0$ such that $\pi_{X_0}(x)>0$, i.e. $\mathcal{X}_0$ is the support of $\mathcal{X}_0$ and let $N_1$ be the same for $X_1$. If $N_0$ or $N_1$ are infinite, the probability distributions are discretized such that these quantities are still large but finite. Let $f(X_0,X_1;\omega)$ be defined as the weighted sum of effects in the parent node of a binary split of the decomposition tree, i.e.:
\begin{gather*}
f(X_0,X_1;\omega)=\sqrt{1-\omega}f_0(X_0)+\sqrt{\omega}f_1(X_1) \qquad \omega \in [0,1]
\end{gather*}
\begin{align*}
f_0(X_0)&=\boldsymbol{D}^T_0(X_0)\boldsymbol{u}_0\qquad \underset{K_0\times 1}{\boldsymbol{u}_0}\sim N(\boldsymbol{0},\boldsymbol{Q}_0^*)\\
f_1(X_1)&=\boldsymbol{D}^T_1(X_1)\boldsymbol{u}_1\qquad
\underset{K_1\times 1}{\boldsymbol{u}_1}\sim N(\boldsymbol{0},\boldsymbol{Q}_1^*).
\end{align*}
Let $\boldsymbol{D}_0(\cdot)$ be a basis made up by $K_0\leq N_0$ functions and $\boldsymbol{D}_1(\cdot)$ by $K_1\leq N_1$ functions.
Let $\boldsymbol{x}_0,\boldsymbol{x}_1$ be row vectors of dimension $N\leq N_0\times N_1$ such that their $i^{th}$ elements form a unique pair $x_{i0}\in\mathcal{X}_0,x_{i1}\in \mathcal{X}_1,\pi_{X_0,X_1}(x_{i0},x_{i1})>0$: the pairs $x_{i0},x_{i1},\;i=1,...,N$ are by design all possible realizations of the covariates that can be observed. Hence, $\boldsymbol{D}_0(\boldsymbol{x}_0)$ and $\boldsymbol{D}_1(\boldsymbol{x}_1)$ become respectively matrices of dimension $K_0\times N$ and $K_1\times N$.

The covariance matrix $\boldsymbol{\Sigma}(\omega)$ of $f(\boldsymbol{x}_0,\boldsymbol{x}_1;\omega)$ is equal to:
\begin{gather*}
\boldsymbol{\Sigma}(\omega)=(1-\omega)\boldsymbol{\Sigma}_0+\omega \boldsymbol{\Sigma}_1\\
\boldsymbol{\Sigma}_0=\boldsymbol{D}^T_0(\boldsymbol{x}_0)\boldsymbol{Q}_0^* \boldsymbol{D}_0(\boldsymbol{x}_0);\qquad
\boldsymbol{\Sigma}_1= \boldsymbol{D}^T_1(\boldsymbol{x}_1)\boldsymbol{Q}_1^* \boldsymbol{D}_1(\boldsymbol{x}_1).
\end{gather*}
An upper bound can be found for the rank of $\boldsymbol{\Sigma}(\omega)$ (called $R(\omega)$ from now on) as $\text{rank}(\boldsymbol{\Sigma}_0)+\text{rank}(\boldsymbol{\Sigma}_1)$. Noting that $R(0)=\text{rank}(\boldsymbol{\Sigma}_0)$ and $R(1)=\text{rank}(\boldsymbol{\Sigma}_1)$ and that the upper bounds for these quantities are:
\begin{equation}\label{eq:R_upper_bounds}
R(0)\leq \min[N_0,K_0]\quad \quad \quad \quad \quad \quad 
R(1)\leq \min[N_1,K_1]
\end{equation}
we can find that an upper bound for $R(\omega)$ is:
\begin{equation}\label{eq:upper_bound}
R(\omega)\leq\min[N_0,K_0]+\min[N_1,K_1].
\end{equation}

Since it might be the case that this upper bound is less than $N$, the probability density $\pi(\boldsymbol{y};\omega)$ of $f(\boldsymbol{x}_0,\boldsymbol{x}_1;\omega)$ can be written using the improper version \citep{RH}, which simplifies to the classical one when $R(\omega)=N$:
\begin{gather}\label{eq:improper_density}
\pi(\boldsymbol{y};\omega)=\dfrac{1}{\sqrt{(2\pi)^{R(\omega)}\cdot|\boldsymbol{\Sigma}(\omega)|^*}}\exp\left(-\frac{1}{2}\boldsymbol{y}^T\boldsymbol{\Sigma}^*(\omega)\boldsymbol{y}\right).
\end{gather}
Note that $|\cdot|^*$ represents the generalized determinant (i.e. the product of non-null eigenvalues).

The first step in deriving the PC prior for $\omega$ is computing the KLD between $f(\boldsymbol{x}_0,\boldsymbol{x}_1;\omega)$ and $f(\boldsymbol{x}_0,\boldsymbol{x}_1;\omega_0)$, where $\omega_0$ is the chosen base model. In this context, we only consider the case $\omega_0=0$; note that the case $\omega_0=1$ is equivalent. 
Once the KLD has been computed as a function of $\omega$, the functional form of the PC prior is found assuming a (truncated) Exponential distribution on the distance $d(\omega;\omega_0)= \sqrt{2\cdot KLD[\pi(\boldsymbol{y};\omega)||\pi(\boldsymbol{y};\omega_0)]}$ and solving for $\omega$.

Using the density function of Equation \eqref{eq:improper_density} for $f(\boldsymbol{x}_0,\boldsymbol{x}_1;\omega)$, the KLD-based distance $d(\omega;\omega_0)$ simplifies to:
\begin{gather}\label{eq:kld_dist}
  d(\omega;\omega_0)=\sqrt{\text{tr}\left[\boldsymbol{\Sigma}^*(\omega_0)\boldsymbol{\Sigma}(\omega)\right]-R(\omega)-\log\frac{|\boldsymbol{\Sigma}(\omega)|}{|\boldsymbol{\Sigma}(\omega_0)|}+[R(\omega_0)-R(\omega)]\log(2\pi)}.
\end{gather}

The KLD-based distance between two multivariate Gaussian distributions simplifies to Equation \eqref{eq:kld_dist} through the following steps:

\begin{align*}
    d(\omega;\omega_0)&=\sqrt{2\cdot KLD[\pi(\boldsymbol{y};\omega)||\pi(\boldsymbol{y};\omega_0)]}\\
    &=\sqrt{2\int \log \left[ \frac{\sqrt{(2\pi)^{R(\omega_0)}\cdot|\boldsymbol{\Sigma}(\omega_0)|}\exp\left(-\frac{1}{2}\boldsymbol{y}^T\boldsymbol{\Sigma}^*(\omega)\boldsymbol{y}\right)}{   
   \sqrt{(2\pi)^{R(\omega)}\cdot|\boldsymbol{\Sigma}(\omega)|} \exp\left(-\frac{1}{2}\boldsymbol{y}^T\boldsymbol{\Sigma}^*(\omega_0)\boldsymbol{y}\right)}
    \right]\pi(\boldsymbol{y};\omega)\text{d} \boldsymbol{y}}\\
    &=\sqrt{[R(\omega_0)-R(\omega)]\log(2\pi)+\log\dfrac{|\boldsymbol{\Sigma}(\omega_0)|}{|\boldsymbol{\Sigma}(\omega)|}+\int \boldsymbol{y}^T[
    \boldsymbol{\Sigma}^*(\omega_0)- \boldsymbol{\Sigma}^*(\omega)] \boldsymbol{y}\cdot \pi(\boldsymbol{y};\omega)\text{d} \boldsymbol{y}}\\
     &=\sqrt{[R(\omega_0)-R(\omega)]\log(2\pi)+\log\dfrac{|\boldsymbol{\Sigma}(\omega_0)|}{|\boldsymbol{\Sigma}(\omega)|}+E_{\pi(\boldsymbol{y};\omega)}\left[ \boldsymbol{y}^T[
    \boldsymbol{\Sigma}^*(\omega_0)- \boldsymbol{\Sigma}^*(\omega)] \boldsymbol{y}\cdot \pi(\boldsymbol{y};\omega)\right]}
\end{align*}

Recalling the formula for the expectation of a quadratic form, it is found that:
\begin{align*}
E_{\pi(\boldsymbol{y};\omega)}\left[ \boldsymbol{y}^T[
    \boldsymbol{\Sigma}^*(\omega_0)- \boldsymbol{\Sigma}^*(\omega)] \boldsymbol{y}\cdot \pi(\boldsymbol{y};\omega)\right]&=\text{tr}\{[\boldsymbol{\Sigma}^*(\omega_0)- \boldsymbol{\Sigma}^*(\omega)]\boldsymbol{\Sigma}(\omega)\}\\
 &=\text{tr}[\boldsymbol{\Sigma}^*(\omega_0)\boldsymbol{\Sigma}(\omega)]-R(\omega)
\end{align*}

Equation \eqref{eq:kld_dist} is still a complicated formula that does not give a simple functional form for the prior of $\omega$, since it directly depends on the chosen basis and precision matrices of the effects involved.

The sum-of-ranks condition consists in checking whether the sum of the number of non-null eigenvalues of $\boldsymbol{\Sigma}_0$ and $\boldsymbol{\Sigma}_1$, or equivalently the sum of their ranks, is less or equal to the dimension $N$. 
\begin{gather}\label{eq:sor_cond}
R(0)+R(1)\leq N.
\end{gather}
This inequality has many consequences. First, it implies that both $\boldsymbol{\Sigma}_0$ and $\boldsymbol{\Sigma}_1$ are singular, assuming that neither of them can be a zero matrix. Most importantly, it guarantees that  $\boldsymbol{\Sigma}(\omega)$ can be rewritten in a more convenient form. 

Let $\boldsymbol{e}_{0,1},...,\boldsymbol{e}_{0,R(0)}$be the $R(0)$ eigenvectors of $\boldsymbol{\Sigma}_0$  associated with non-null eigenvalues $\lambda_{0,1},...,  \lambda_{0,R(0)}$. Let $\boldsymbol{e}_{1,1},...,\boldsymbol{e}_{1,R(1)}$ be the $R(1)$ eigenvectors of $\boldsymbol{\Sigma}_1$  associated with non-null eigenvalues $\lambda_{1,1},...,  \lambda_{1,R(1)}$. If the sum-of-ranks condition (\ref{eq:sor_cond}) holds, then $\boldsymbol{\Sigma}(\omega)$ can be rewritten in terms of $\boldsymbol{\Lambda}_0=\text{diag}(\lambda_{0,1},...,  \lambda_{0,R(0)},0....,0)$, $\boldsymbol{\Lambda}_1=\text{diag}(0,...,0,\lambda_{1,1},...,  \lambda_{1,R(1)})$ and a common eigenbasis $\boldsymbol{V}=[\boldsymbol{e}_{0,1},...,\boldsymbol{e}_{0,R(0)},\boldsymbol{0},...,\boldsymbol{0},\boldsymbol{e}_1,...,\boldsymbol{e}_{R(1)}]$:
\begin{gather*}
\boldsymbol{\Sigma}(\omega)=\boldsymbol{V}[(1-\omega)\boldsymbol{\Lambda}_0+\omega\boldsymbol{\Lambda}_1]\boldsymbol{V}^T.
\end{gather*}

Rewriting $\boldsymbol{\Sigma}(\omega)$ ensures that the distance function (\ref{eq:kld_dist}) can be further simplified.

Under the assumption of a common eigenbasis $\boldsymbol{V}$ between the covariance matrices $\boldsymbol{\Sigma}_0$ and $\boldsymbol{\Sigma}_1$, it is possible to simplify the distance following the proof provided in \cite{FV22}. First, note that $\boldsymbol{\Sigma}(\omega)$ and its generalized inverse can be rewritten as:
\begin{align*}
\boldsymbol{\Sigma}(\omega)&=\boldsymbol{V}\left[(1-\omega)\boldsymbol{\Lambda}_0+\omega\boldsymbol{\Lambda}_1\right]\boldsymbol{V}^T\\
\boldsymbol{\Sigma}^*(\omega)&=\boldsymbol{V}\left[(1-\omega)\boldsymbol{\Lambda}_0+\omega\boldsymbol{\Lambda}_1\right]^*\boldsymbol{V}^T\\
\end{align*}

Then, it can be noticed that the first term of Equation \eqref{eq:kld_dist} simplifies and becomes only a function of the eigenvalues in $\boldsymbol{\Lambda}_0$ and $\boldsymbol{\Lambda}_1$:
\begin{align*}
\text{tr}\left[\boldsymbol{\Sigma}^*(\omega_0)\boldsymbol{\Sigma}(\omega)\right]&=\text{tr}[\boldsymbol{V}\left[(1-\omega_0)\boldsymbol{\Lambda}_0+\omega_0\boldsymbol{\Lambda}_1\right]^*\boldsymbol{V}^T \boldsymbol{V}\left[(1-\omega)\boldsymbol{\Lambda}_0+\omega\boldsymbol{\Lambda}_1\right]\boldsymbol{V}^T]\\
&=\text{tr}[\left[(1-\omega_0)\boldsymbol{\Lambda}_0+\omega_0\boldsymbol{\Lambda}_1\right]^*\left[(1-\omega)\boldsymbol{\Lambda}_0+\omega\boldsymbol{\Lambda}_1\right]]\\
&=\sum_{n:\lambda_{0,n}+\lambda_{1,n}> 0}\frac{(1-\omega)\lambda_{0,n}+\omega\lambda_{1,n}}{(1-\omega_0)\lambda_{0,n}+\omega_0\lambda_{1,n}}
\end{align*}
Finally, also the second term of Equation \eqref{eq:kld_dist} can be rewritten simly in terms of eigenvalues under the common eigenbasis assumption:
\begin{align*}
\log\frac{|\boldsymbol{\Sigma}(\omega)|}{|\boldsymbol{\Sigma}(\omega_0)|}&=\log\frac{\left|(1-\omega)\boldsymbol{\Lambda}_0+\omega\boldsymbol{\Lambda}_1\right|}{\left|(1-\omega_0)\boldsymbol{\Lambda}_0+\omega_0\boldsymbol{\Lambda}_1\right|}\\
&=\sum_{n:\lambda_{0,n}+\lambda_{1,n}> 0}\log\frac{(1-\omega)\lambda_{0,n}+\omega\lambda_{1,n}}{(1-\omega_0)\lambda_{0,n}+\omega_0\lambda_{1,n}}
\end{align*}

At this point, the distance function can be further simplified because under the sum-of-ranks condition, for each $n\in[1,N]$, at least one eigenvalue between $\lambda_{0,n}$ and $\lambda_{1,n}$
will be 0. This leads to the following simplification:
\begin{gather*}
    \frac{(1-\omega)\lambda_{0,n}+\omega\lambda_{1,n}}{(1-\omega_0)\lambda_{0,n}+\omega_0\lambda_{1,n}}=\begin{cases}
       \frac{1-\omega}{1-\omega_0},\;\;\;\lambda_{0,n}>0,\lambda_{1,n}=0\\
        \frac{\omega}{\omega_0},\;\;\;\;\;\lambda_{0,n}=0,\lambda_{1,n}>0      
    \end{cases}
    \end{gather*}
    
The final form of the distance function is the following:
\begin{equation*}
    d(\omega;\omega_0)=\sqrt{
    \begin{aligned}
     &R(0)\frac{1-\omega}{1-\omega_0}+R(1)\frac{\omega}{\omega_0}-R(0)\log\frac{1-\omega}{1-\omega_0}-R(1)\log\frac{\omega}{\omega_0}\\
     &-R(\omega)+[R(\omega_0)-R(\omega)]\log(2\pi)
    \end{aligned}
    }
\end{equation*}

The function does not include the eigenvalues of $\boldsymbol{\Sigma}_0$ and $\boldsymbol{\Sigma}_1$ but it still depends on $R(\omega)$, which is computationally inefficient. However, noting that the distance is not finite for $\omega_0=0$, it is necessary to compute it as a limit. Consider first $d(\omega;\omega_0)^2\omega_0$:
\begin{align*}
d(\omega;\omega_0)^2\omega_0&=\omega_0 R(0)\frac{1-\omega}{1-\omega_0}-\omega_0R(0)\log\frac{1-\omega}{1-\omega_0}\\
&+R(1)\omega-\omega_0 R(1)\log\frac{\omega}{\omega_0}\\
&-\omega_0 R(\omega)+\omega_0[R(\omega_0)-R(\omega)]\log(2\pi)   
\end{align*}

Computing the limit for $\omega_0\rightarrow 0$, it is found that:
\begin{gather*}
   \underset{\omega_0\rightarrow 0}{lim} d(\omega;\omega_0)^2\omega_0=R(1)\omega   
\end{gather*}

However, this distance is not finite for $\omega_0=0$ and must be computed instead as a limit for $\omega_0=0$ \citep{FV22}. It can be proven that:
\begin{gather}\label{eq:simple_dist}
   \underset{\omega_0\rightarrow 0}{\lim} d(\omega;\omega_0)=R(1)\sqrt{\omega}  
   \end{gather}
   where $R(1)$ is a constant with respect to $\omega$.
Specifying an Exponential distribution on the distance (truncated at $d(1)$ as the upper bound for $\omega$ is 1), the final form of the PC prior is found using the change-of-variable formula:
\begin{gather*}
   \pi(\omega)=\frac{\widetilde{\delta} R(1) \exp(-\widetilde{\delta} R(1) \sqrt{\omega})}{2\sqrt{\omega}[1-\exp(-\widetilde{\delta} R(1))]}.
\end{gather*}
The dependence on the constant $R(1)$, which contains parameters of the model is removed, as it becomes not identifiable with the hyperparameter of the Exponential distribution $\widetilde{\delta}$. Denoting $\delta=\widetilde{\delta}R(1)$, we obtain Equation \eqref{eq:simplified_form}, which proves that, under the sum-of-ranks condition, the simplified form of the $\text{PC}_0$ prior is guaranteed.

The $\delta$ hyperparameter can be chosen using probability statements such as $P(\omega < U)=\alpha$. However, in order to obtain a valid probability distribution (i.e. $\delta>0$), \cite{FV22} noted that the following condition must be respected:
\begin{gather}\label{eq:U_bound}
\alpha\geq \sqrt{U}.
\end{gather}
As a consequence, the median can be at most $0.25$, which can be obtained with $\delta \rightarrow 0$.

Because the sum-of-ranks condition guarantees a simpler solution for the PC prior, it is convenient to start the derivation of the prior by checking this condition first.
If the condition is not found to be satisfied using the upper bounds of Equation \eqref{eq:R_upper_bounds}, the next step should be to derive the actual $R(0)$ and $R(1)$ and check the condition again. 

The sum-of-ranks assumption can be verified when $R(0)$ and $R(1)$ are available analytically, but this is rarely the case in practice. Therefore, the assumption can be checked instead using their upper bounds (Equation \eqref{eq:R_upper_bounds}), which are usually explicitly available to  the user. We discuss here three cases in which $\text{PC}_0$ priors on $\omega$ are a sensible choice and may result from the decomposition tree designed for SDMs (Figure \ref{fig:default_tree}).

\paragraph{Linear effect versus non-linear effect.}
Let $X=X_0=X_1$ be a univariate variable. Hence, $N_0=N_1=N$ since the two covariates are completely dependent. Let $f_0(X)$ be a linear effect (after standardization of $X$) and $f_1(X)$ a non-linear one with a finite number of $K_1$ coefficients, constrained by design to have a null intercept and linear trend. The upper bounds for $R(0)$ and $R(1)$ are: $R(0)\leq 1,\;
R(1)\leq K_1$
As a consequence, their sum $R(0)+
R(1)\leq K_1+1$ respects the condition of Equation \eqref{eq:sor_cond} as long as $K_1<N$. This scenario can emerge from splits at Level 4 of the tree from Figure \ref{fig:default_tree}.

\paragraph{Linear main effects versus  interaction effect.}

Let $X_0=X_1=[X_A,X_B]$ where $X_A,X_B$ are two independent univariate covariates. Again, $N_0=N_1=N$ and $N=N_A\cdot N_B$.
Let $f_0(X_A,X_B)$ and $f_1(X_A,X_B)$ be defined as:
\begin{gather*}
f_0(X_A,X_B)=\sqrt{1-\phi}\widetilde{X_A} u_A+\sqrt{\phi} \widetilde{X_B}u_B;\qquad \quad
f_1(X_A,X_B)=\widetilde{X_A}\widetilde{X_B}u_1
\end{gather*}
where $\widetilde{X_A},\widetilde{X_B}$ are the standardized versions of $X_A,X_B$. In this case, $R(1)\leq 1$.  $R(0)$ depends on the value of $\phi$ but its upper bound can still be found using the same formula of Equation \eqref{eq:upper_bound}: $
R(0)\leq \text{min}[N_A,1]+\text{min}[N_B,1]\implies 
R(0)\leq 2.$
Hence, the sum-of-ranks condition is respected as long as $3\leq N_A\cdot N_B$, which is always true if $X_A,X_B$ are both continuous covariates. This case covers the splits at Level 3 in the left branch of the tree from Figure \ref{fig:default_tree}.

\paragraph{Finite-dimensional main effects versus Kronecker product interaction effect.}

Consider the scenario above where now $f_0(X_A,X_B)$ is defined as the weighted sum of two finite-dimensional effects for $X_A$ and $X_B$ with respectively $K_A$ and $K_B$ coefficients with $K_A\leq N_A, K_B\leq N_B$:
\begin{gather*}
f_0(X_A,X_B)=\sqrt{1-\phi}\boldsymbol{D}_A^T(X_A)\boldsymbol{u}_A+\sqrt{\phi} \boldsymbol{D}_B^T(X_B)\boldsymbol{u}_B\\
\underset{K_A\times 1}{\boldsymbol{u}_A }\sim N(\boldsymbol{0},\boldsymbol{Q}_A^*);\qquad\qquad
\underset{K_B\times 1}{\boldsymbol{u}_B }\sim N(\boldsymbol{0},\boldsymbol{Q}_B^*).
\end{gather*}
Let $f_1(X_A,X_B)=\boldsymbol{D}^T(X_A,X_B)\boldsymbol{u}$ be an interaction effect built using the Kronecker product between the basis and precision matrices of the main effects in $f_0(X_A,X_B)$, i.e. the $IV$ type of interaction effect defined by \cite{knorr2000bayesian} for inseparable spatio-temporal effects.
\begin{align*}
\boldsymbol{D}(X_A , X_B)&=\boldsymbol{D}_A (X_A) \otimes \boldsymbol{D}_B (X_B);\qquad
\underset{K_A\cdot K_B \times 1}{\boldsymbol{u}} &\sim N(\boldsymbol{0},[\underset{K_A\times K_A}{\boldsymbol{Q}_A} \otimes \underset{K_B\times K_B}{\boldsymbol{Q}_B}]^*).
\end{align*}
The upper bounds for $R(0)$ and $R(1)$ can be easily found to be $
R(0)\leq K_A+K_B\qquad
R(1) \leq K_A\cdot K_B$.
Hence, the sum-of-ranks condition is satisfied as long as: $
K_A+K_B+ K_A\cdot K_B\leq N_A\cdot N_B$.
This is always true if $X_A$ and $X_B$ are continuous covariates. The condition also holds in many other cases in which $X_A$ and $X_B$ are discrete, such as the particular example derived in \cite{FV22}. 
This last case covers the possible split between the main spatial and temporal effects and their non-additive interaction (Level 3, right branch of Figure \ref{fig:default_tree}).

\end{document}